# In-operando quantitation of Li concentration for commercial Li-ion rechargeable battery using high-energy X-ray Compton scattering

Kosuke Suzuki[a]*, Ayahito Suzuki[a], Taiki Ishikawa[a], Masayoshi Itou[b], Hisao Yamashige[c], Yuki Orikasa[d], Yoshiharu Uchimoto[e], Yoshiharu Sakurai[b] and Hiroshi Sakurai[a]

[a] Department of Electronics and Informatics, Gunma University, 1-5-1, Tenjin-cho, Kiryu, Gunma, 376-8515, Japan

[b] Japan Synchrotron Radiation Research Institute (JASRI), 1-1-1 Kouto, Sayo, Hyogo, 679-5198, Japan

[c] Material Platform Engineering Division, TOYOTA MOTOR CORPORATION, 1 toyota-cho, Toyota, Aichi, 471-8572, Japan

[d] Department of Applied Chemistry, Ritsumeikan University, 1-1-1 Noji-higashi, Kusatsu, Shiga, 525-8577, Japan

[e] Guraduate Shcool of Human and Environmental Studies, Kyoto University, Sakyo, Kyoto, 606-8501, Japan

Correspondence email: kosuzuki@gunma-u.ac.jp

**Abstract**   Compton scattering is one of the promising probe to quantitate of the Li under *in-operando* condition, since high-energy X-rays which have high penetration power into the materials are used as incident beam and Compton scattered energy spectrum have specific line-shape by the elements. We develop *in-operando* quantitation method of Li composition in the electrodes by using line-shape (*S*-parameter) analysis of Compton scattered energy spectrum. In this study, we apply *S*-parameter analysis to commercial coin cell Li-ion rechargeable battery and obtain the variation of *S*-parameters during charge/discharge cycle at positive and negative electrodes. By using calibration curves for Li composition in the electrodes, we determine the change of Li composition of positive and negative electrodes through *S*-parameters, simultaneously.

## 1. Introduction

*In-operando* quantitation of Li ions which contribute to electrodes reaction is crucially important to produce high performance Li-ion rechargeable battery equipped high efficiency, high safety, and long lifetime. Generally, the electrodes of Li-ion battery are made from several compounds, active material, conductive aid, and binder. With such electrodes, therefore, it has been reported that inhomogeneous electrode reaction occurs (Nishi et al. 2013). This inhomogeneous reaction induces an overvoltage by repeating charge/discharge cycle. It is important to monitor Li ion distributions in the electrodes in order to understand where the overvoltage state is made. Although development of experimental technique to visualize Li distribution have attracted much attention, these experimental methods are mainly adopted to only positive electrode (Liu et al., 2010; Mima et al., 2012; Wang et al., 2014; Jensen et al., 2015; He et al., 2015). The reactions in whole of the battery have been observed by using neutron source. However, the spatial resolution is limited as compared with X-ray source (Wang et al., 2012).

So far, we have been investigating the mechanism of electrode reactions in positive electrode materials, $Li_xMn_2O_4$, $Li_xCoO_2$ and $Li_xFePO_4$ (Suzuki et al., 2015; Barbiellini et al., 2016; Hafiz et al., in submitted) and developing *in-operando* quantitation method of Li ions by using high-energy X-ray Compton scattering spectroscopy. The merits of Compton scattering technique are that Compton scattering experiment yield bulk observations on working batteries since the X-ray photons reaching over 100 keV were used as incident beam, and these photons can penetrate deep into materials, including closed electrochemical cells and Compton scattered energy spectrum, so-called Compton profile, have specific line-shape by each element. In previous study, we developed how to quantify Li concentration by introducing shape parameter (*S*-parameter) of Compton profile. In a Compton profile, the contribution of Li concentration clearly appears. The *S*-parameter is a parameter to digitalize the line-shape of Compton profile. By using *S*-parameter, we confirmed that the experimental *S*-parameters obtained from polycrystalline $Li_xMn_2O_4$ pellets are proportional to Li composition of the samples obtained from inductively coupled plasma (ICP) analysis. Moreover, we successfully observed non-destructively the change of Li concentration of commercial CR2032 coin battery through the *S*-parameters, (Suzuki et al., 2016). However, this Li concentration was obtained from only positive electrode. Hence, in this study, we apply *S*-parameter analysis to commercial Li-ion secondary battery, VL2020, and observe Li concentration including positive and negative electrodes with charge/discharge cycle.

## 2. Experiment and data analysis

A coin cell Li-ion rechargeable battery VL2020 which made from Panasonic Corporation was used for Compton profile measurement. The VL2020 is composed $V_2O_5$ positive electrode with 800 μm thickness, LiAl alloy negative electrode with about 300 μm thickness, olefin-based nonwoven fabric separator, Al wire-netting spacer, and dimethoxyethane electrolyte as shown in figure 1. Compton scattering experiments were performed on the BL08W beamline at SPring-8 synchrotron facility, Japan. Fig. 1 shows experimental setup of Compton scattering experiment. Synchrotron radiation emitted from elliptical multipole wiggler was used as incident X-rays. The incident X-ray energy tuned to 115 keV by asymmetric Johann-type Si (400) monochrometer. The incident X-rays were delivered to coin battery after formed to 25 μm height and 500 μm width by incident slit. Compton scattered X-ray intensities were measured by using 9 segments Ge solid-state detector (SSD). A collimator slit with 500 μm diameter is arranged between the sample and the detector. The scattering angle was fixed 90 degrees. The probing volume of measurement is decided by incident and collimator slits, the size is 25 μm × 500 μm × 500 μm. The coin battery was connected to charge/discharge device (HZ7000, Hokuto Denko Corporation) and charge state of the battery was changed by charging or discharging at 0.4C rate. The Li concentrations for full charged state (state of charge: SOC100) and full discharged state (SOC0) of positive and negative electrodes were confirmed by ICP analysis in the Center for Instrumental Analysis of Gunma University.

By measuring Compton scattered X-ray energy spectrum, we can obtain Compton profile. The Compton profile, $J(p_z)$, defines following equation (Cooper et al., 2004),

$$J(p_z) = \iint \rho(\mathbf{p}) dp_x dp_y, \qquad (1)$$

where $\rho(\mathbf{p})$ is electron momentum density, $\mathbf{p} = (p_x, p_y, p_z)$ is momentum and $p_z$ is taken to lie along the direction of the scattering vector. The $\rho(\mathbf{p})$ is written by following equation (Barbiellini 2001; Barbiellini et al., 2002),

$$\rho(\mathbf{r}) = \sum_j n_j \left| \int \Psi_j(\mathbf{r}) \exp(-i\mathbf{p} \cdot \mathbf{r}) dr \right|^2, \qquad (2)$$

where $\Psi_j(\mathbf{r})$ the wave function of electron in the $j$-state and $n_j$ is the corresponding electron occupation. The index j runs over all constituent atoms and orbitals. The line-shape of a Compton profile varies depending on each element because the Compton profile directly links to wave-function of the electron. The $S$-parameter is defined as eq. (3),

$$S = \frac{S_L}{S_H} = \frac{\int_{-d}^{d} J(p_z) dp_z}{\int_{-r}^{-d} J(p_z) dp_z + \int_{d}^{r} J(p_z) dp_z}, \qquad (3)$$

where $S_L$ and $S_H$ are the areas under the Compton profile covering the low and high momentum regions. The parameters $d$ and $r$ are the range of the low momentum and high momentum regions, respectively. In this study, $d = 1$ atomic units (a.u.) and $r = 5$ a.u. are used. The detail of decision method for parameter $d$ and $r$ is given in the Supporting information.

## 3. Results and discussion

Figure 2(a) shows energy spectra for stainless steel (SUS), positive electrode, separator, and negative electrode. In order to compare the line-shape of these energy spectra, the area of each spectrum is normalized to same value. In Fig. 2(a), the energy spectrum for separator has sharpest line-shape and the energy spectrum for SUS has broadest line-shape. Therefore, $S$-parameter obtained from separator energy spectrum has highest value. On the other hand, $S$-parameter obtained from SUS energy spectrum has smallest value. $S$-parameters which obtained scanning incident X-rays along vertical ($z$) direction of the battery are shown in Fig. 2(b). In this figure, the region of $z < 0$ mm and $1.5$ mm $< z$ correspond to battery outer case (SUS), the region of $0$ mm $< z < 0.15$ mm corresponds to spacer (Al wire-netting), the region of $0.15$ mm $< z < 0.4$ mm corresponds to negative electrode (LiAl), the region of $0.4$ mm $< z < 0.65$ mm corresponds to separator, and the region of $0.65$ mm $< z < 1.5$ mm corresponds to positive electrode ($V_2O_5$), respectively. The structure in the VL2020 battery is revealed through $S$-parameters.

We measured Compton scattered X-ray energy spectrum by scanning incident X-rays along $z$ direction on the battery with changing the state of charge (SOC). The charge-discharge curve in this measurement is shown in Fig. 3(a). Initial voltage is 3.507 V. Full charged state voltage is 3.5 V and full discharged state voltage is 2.5 V. The SOC is changed every 2.5 hours by constant current of 8 mA. There is rest time of 30 minutes between 1st-discharge and 1st-charge and between 1st-charge and 2nd-discharge. After 2nd-discharge, battery reactions were relaxed for 3 hours. The z position scan with 66 steps was repeated 34 times for 11.6 hours to obtain $S$-parameter distribution. Measurement time is 15 seconds at one data point. The z of y-axis corresponds to vertical position of the battery as shown in Fig. 3(b). $S$-parameter distribution obtained from Compton scattered X-ray energy spectrum is shown in Fig. 3(c). Colour distribution corresponds to $S$-parameter intensity. Variation of $S$-parameters appears at upper area of positive electrode (around $z = 1.22$ mm) and at the interface between separator and negative electrode (around $z = 0.3$ mm). Moreover, we observed that the separator position moves toward to positive electrode during charge and returns to initial position during discharge. It is thought that this migration of separator position is induced by expansion or contraction of lattice volume in the negative electrode. Because the volume expansion from Al to LiAl is 95 % (Morales et al., 2010), on the other hand, the volume expansion from $V_2O_5$ to $Li_{0.5}V_2O_5$ is 2.2% (Voikov et al., 1994; Wachter et al., 2014).

In order to clear the change of $S$-parameters during charge/discharge cycle, the separator position was corrected. The correction which assumed volume expansion in negative electrode was performed by using following equation,

$$z_c = az_i^2 + bz_i, \tag{4}$$

Here, $z_c$ and $z_i$ are corrected and initial z position of the battery, respectively. The coefficients *a* and *b* are determined by least square method.

Fig. 4 shows difference *S*-parameters ($\Delta S$) distribution during charge/discharge process. $\Delta S$ is defined as equation (5),

$$\Delta S = S_t - S_0, \tag{5}$$

here, $S_t$ is *S*-parameters at *t* seconds, $S_0$ is averaged *S*-parameter of the whole measurement time in each z position. In Fig. 4, it seems that the *S*-parameters at positive electrode increase during discharge and decrease during charge homogeneously. On the other hand, variation of *S*-parameters at negative electrode mostly appears at the interface between negative electrode and separator. In previous study, we confirmed that there is linear relation between *S*-parameters and Li concentration (Suzuki et al., 2016). Hence, we can think that the variation of *S*-parameter in positive and negative electrodes during charge/discharge cycle correspond to the change of Li concentration. It indicates that inhomogeneous lithium reactions easily occur in negative electrode. We observe the change of *S*-parameters at separator region as well. In order to consider this change of *S*-parameters at separator region, raw *S*-parameter data at separator region are shown in Fig. 5. By charging the battery, value of *S*-parameters increases then the value returns to initial value by discharge. It indicates that some Li ions remain into separator when Li ions move from positive electrode to negative electrode by charge cycle. As one possible reason, this residue of Li ions induces capacity loss of the battery.

Fig. 6 shows the average of *S*-parameters during charge/discharge at whole of positive and negative electrodes. The averaged *S*-parameters variation are obtained from *S*-parameters of every one scan at positive and negative electrodes regions in Fig. 3(c). The *S*-parameters decrease 0.92 % in positive electrode from discharge to charge and increase 1.69 % in negative electrode from discharge to charge cycle, respectively. In order to quantitate of Li concentration, *x*, from *S*-parameters, theoretical *S*-parameters of $Li_xV_2O_5$ (*x* = 0, 0.124, 0.25, 0.5, 0.75, and 1) and $Li_xAl$ (*x* = 0, 0.046, 0.25, 0.5, 0.75, and 1) were calculated by using atomic model Compton profile data (Biggs et al., 1975). As a reference data of Li composition, ICP analysis were performed for positive and negative electrodes in SOC100 and SOC0, respectively. The mass of Li at SOC100 is 1.497 mg/l in positive electrode and 5.068 mg/l in negative electrode and at SOC0 is 5.373 mg/l in positive electrode and 1.211 mg/l in negative electrode, respectively. The molar ratio of Li, *x*, is calculated by using following equations,

$$x = \frac{M_{VO} \cdot m_{Li}}{M_{Li} \cdot m_{LVO}} \text{ (in positive electrode)}, \quad x = \frac{M_A \cdot m_{Li}}{M_{Li} \cdot m_{LA}} \text{ (in negative electrode)}, \tag{6}$$

here, $M_{VO}$ (= 181.878) is the molecular weight for $V_2O_5$. $M_A$ (= 26.982) and $M_{Li}$ (= 6.941) are the atomic weight for Al, and Li atoms, respectively. The $m_{LVO}$, $m_{LA}$, and $m_{Li}$ are the mass of $Li_xV_2O_5$, $Li_xAl$, and Li, respectively. The $m_{LVO}$ and $m_{LA}$ of SOC100 and SOC0 were measured by using the electronic balance. The values of $Li_xV_2O_5$ are 330.7 mg in SOC0 and 317.4 mg in SOC100 and of $Li_xAl$ are 103.0 mg in SOC0 and 110.6 mg in SOC100, respectively. The values obtained from ICP analysis were used as $m_{Li}$. The obtained lithium composition of $Li_xV_2O_5$ is 0.426 in SOC0 and 0.124 in SOC100 and of $Li_xAl$ is 0.046 in SOC0 and 0.178 in SOC100, respectively. Since momentum resolution and background is different between experimental and theoretical data, the theoretical *S*-parameters of $Li_{0.124}V_2O_5$ is normalized by the experimental value of *S*-parameter in SOC100. Similarly, the theoretical *S*-parameters of $Li_{0.046}Al$ is normalized by the experimental value of *S*-parameter in SOC0. Fig. 7(a) and (b) show the calibration curves for $Li_xV_2O_5$, and $Li_xAl$ and that is $y = 32.609x - 47.186$ in $Li_xV_2O_5$, and $y = 5.299x - 6.967$ in $Li_xAl$, respectively. By using these calibration curves, the Li compositions, $x$, for SOC0 in positive electrode and for SOC100 in negative electrode are obtained as $x = 0.470 \pm 0.012$, and $x = 0.170 \pm 0.006$, respectively. These values are equal with the values obtained from ICP analysis. Fig. 8 shows variation of Li composition in $Li_xV_2O_5$ and $Li_xAl$ during charge/discharge cycle. These Li compositions were obtained by applying calibration curves to the data of Fig. 6. We successfully quantitated Li concentration of positive and negative electrodes in commercial Li-ion rechargeable battery during charge/discharge cycle, simultaneously.

## 4. Summary

In this study, we applied *S*-parameter analysis of Compton scattered energy spectrum to commercial coin cell Li-ion rechargeable battery, VL2020. The structure of the VL2020 was revealed through *S*-parameters. *S*-parameter distribution correspond to charge/discharge cycle is obtained at positive and negative electrodes. When the battery was charged, the migration of the separator position induced by lattice expansion of negative electrode was observed. By using calibration curves obtained from theoretical *S*-parameters, we obtained Li concentration of positive and negative electrodes during charge/discharge cycle, simultaneously.

**Acknowledgements**   The authors thank Mr. K. Iwamaru, Prof. F. Hayashi, and Prof. K. Tsunoda of Gunma University for technical support of ICP measurements. This work was supported by Japan Science and Technology Agency, and Grant-in-Aid for Young Scientists (B) from MEXT KAKENHI under Grant Nos. 24750065 and 15K17873. The Compton scattering experiments were performed with the approval of JASRI [Proposal Nos. 2013A1007, 2013B1019, 2014A1012, 2014B1023, 2015A1010, 2015A1477, 2015B1010, 2015B1213, 2016A1016, 2016A1211, and 2016B1020].

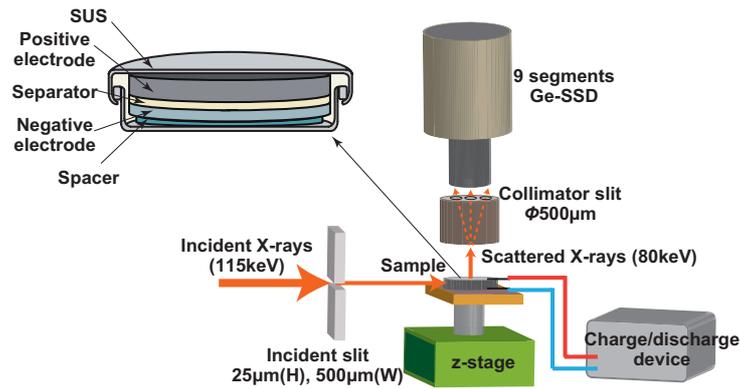

**Figure 1** Experimental setup of Compton scattering experiment at BL08W of SPring-8 and configuration of coin battery of VL2020.

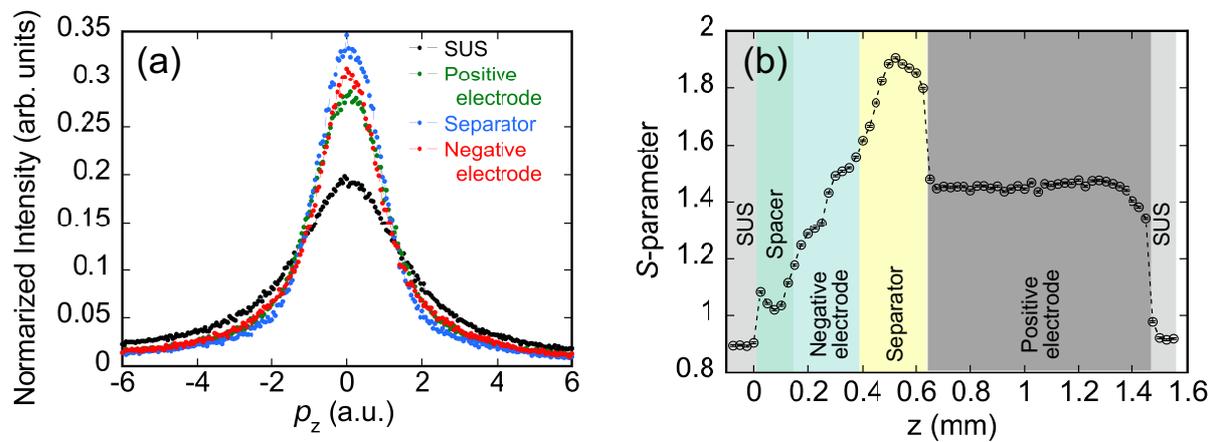

**Figure 2** (a) Compton scattered energy spectrum for SUS (black dots), positive electrode (green dots), separator (blue dots), and negative electrode (red dots). In order to compare the line-shape of spectra, the area of each spectrum is normalized to same value. (b) Battery component through $S$-parameters. The z in x-axis correspond to z direction of the battery.

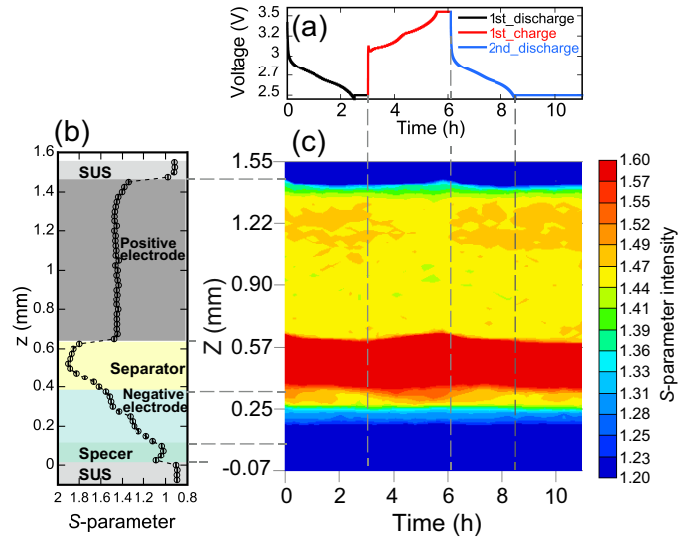

**Figure 3** (a) Charge/discharge curve of VL2020. 1st discharge, 1st charge and 2nd discharge correspond to black, red and blue solid lines. (b) Component of the battery. (c) *S*-parameter distribution during charge/discharge cycle.

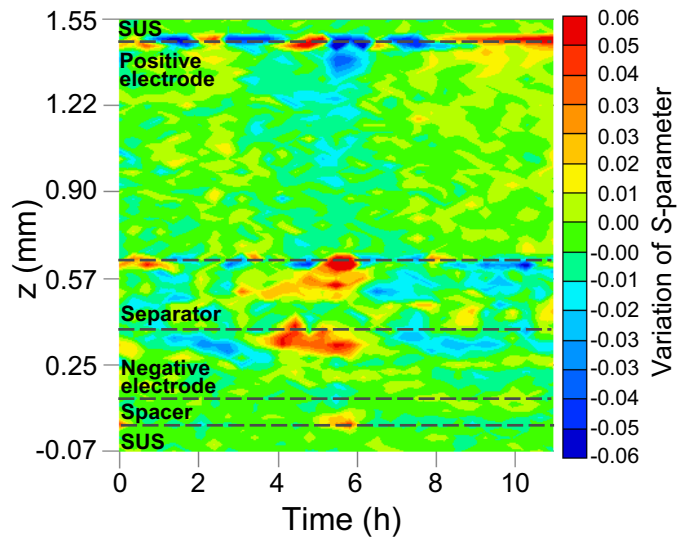

**Figure 4** The variation of *S*-parameter distribution.

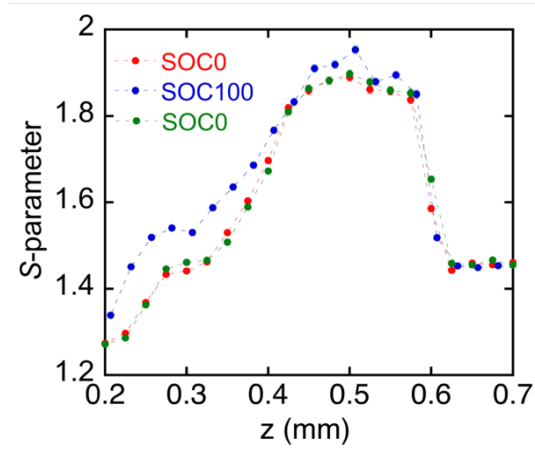

**Figure 5** The raw *S*-parameters data at separator region. Red and green dots data are obtained by 1st discharge and 2nd discharge and blue dots data is obtained by 1st charge of the battery.

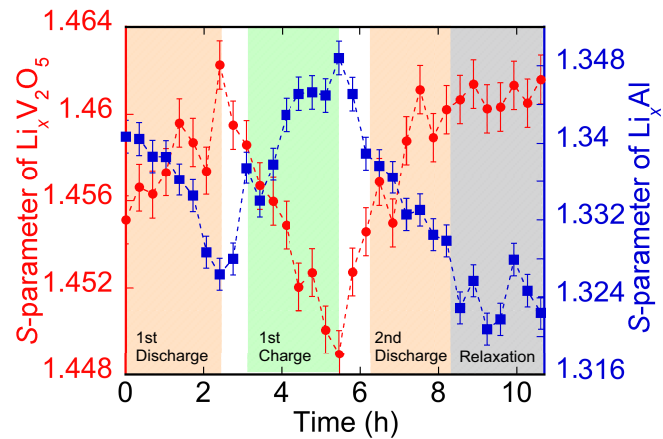

**Figure 6** The variation of *S*-parameters during charge/discharge cycle in $Li_xV_2O_5$ positive electrode material (red dots) and $Li_xAl$ negative electrode material (blue squares). Red background color regions correspond to 1st discharge and 2nd discharge, respectively. Green and gray background color regions correspond to 1st charge and relaxation.

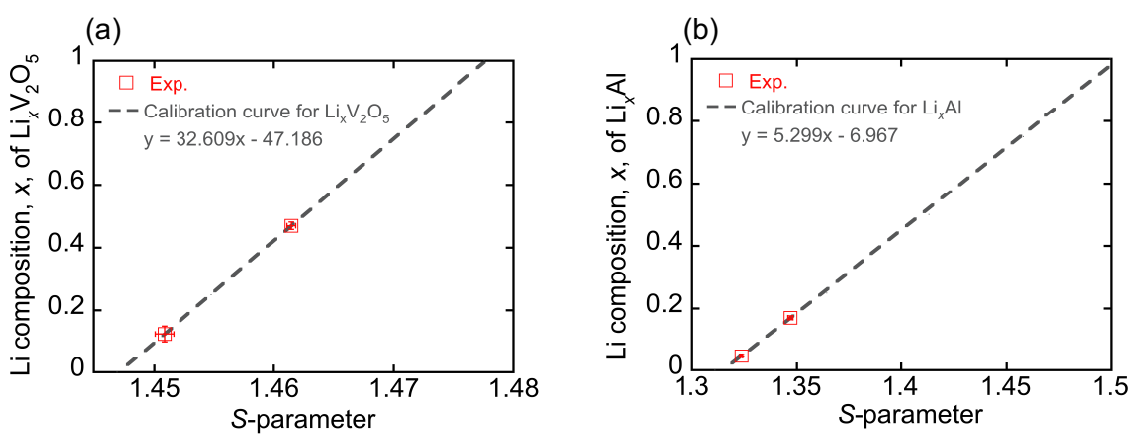

**Figure 7** Calibration curves for Li$_x$V$_2$O$_5$ and Li$_x$Al are calculated by using atomic model Compton profile. The theoretical *S*-parameters for Li$_{0.124}$V$_2$O$_5$ normalized by the value of experimental *S*-parameter of SOC100 in positive electrode. The theoretical *S*-parameter for Li$_{0.046}$Al normalized by the value of experimental *S*-parameter of SOC0 in negative electrode.

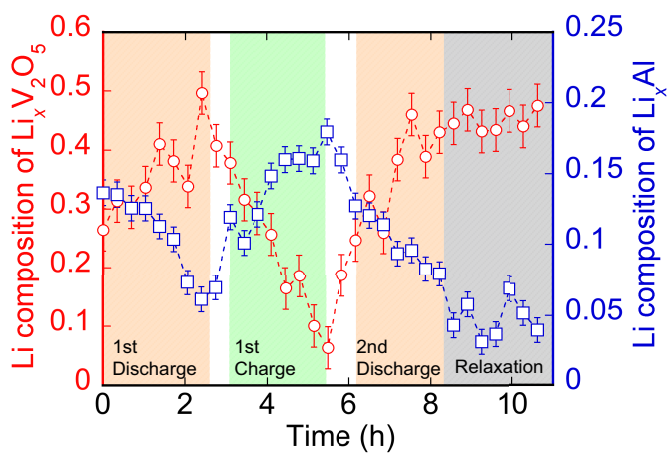

**Figure 8** The change of Li composition obtained from the calibration curve during charge/discharge. Red dots and blue dots correspond to Li concentration in Li$_x$V$_2$O$_5$ positive electrode material and Li$_x$Al negative electrode material, respectively. Red background color regions correspond to 1st discharge and 2nd discharge, respectively. Green and gray background color regions correspond to 1st charge and relaxation.

## Supporting information

### S1. Determination of the boundary for *S*-parameter analysis

In order to determine the parameters of *d* and *r* in *S*-parameter analysis, energy spectra of positive electrode in SOC0 and SOC100 were measured. Figure S1 shows difference energy spectrum, $\Delta I$. Here, $\Delta I$ is defined as $\Delta I = I_{SOC0} - I_{SCO100}$. The $I_{SOC0}$ and $I_{SOC100}$ are energy spectra for full charged state and full discharged state, respectively. In order to reveal the boundary of low momentum and high momentum regions, Gaussian model fitting was performed to difference energy spectrum. In the Fig. S1, we observed small peak which correspond to increment of Li ions around $p_z = 0$ a.u. It is confirmed that the contribution of Li ions appears at low region previously (Suzuki et al., 2016). There is Li contribution from $p_z = -1$ a.u. to $p_z = 1$ a.u. At $p_z = \pm 1.5$ a.u., negative part appears. This negative part indicates the change of electronic structure by Li insertion as it has been found from previous study of $Li_xMn_2O_4$ (Suzuki et al., 2015). At $p_z > 5$ a.u. and $p_z < -5$ a.u., the variation between $I_{SOC0}$ and $I_{SOC100}$ is almost zero. Hence, we use $d = 1$ a.u. as the low momentum region and $r = 5$ a.u. as high momentum region. This boundary was applied to analysis of negative electrode since same trend shows in $Li_xAl$.

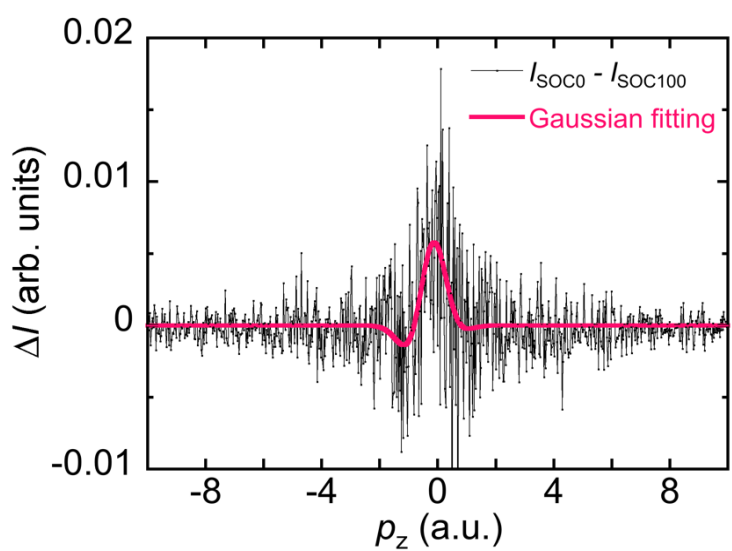

**Figure S1** Difference energy spectrum (black solid line) between the energy spectra of SOC0 and SOC100 in positive electrode materials. Red solid line shows the result of Gaussian model fitting.